\documentclass{article}
\usepackage{hiph-preprint}
\usepackage{graphicx}
\volnumber{22} \issuenumber{1} \edyear{2005}                             
\frompage{000} \topage{000}                                              
\recrevdate{10 November 2005}                                              
\def\rightmark{Jet topologies via three particle correlations}
\def\leftmark{N. N. Ajitanand}

\title{Identification of exotic jet topologies via three particle correlations in PHENIX}
\authors{ 
{N. N. Ajitanand$^1$
\index{One, N. N. Ajitanand,  for the PHENIX Collaboration} 
}\\[2.812mm]
{\normalsize
\hspace*{-8pt}$^1$ Dept. of Chemistry, 
        SUNY Stony Brook, \\ 
Stony Brook, NY 11794, USA\\[0.2ex] 
}}
\abstract{Modifications of jet properties resulting from the coupling of jets to the strongly 
interacting matter produced in RHIC collisions are of great current interest. 
In recent work, the PHENIX collaboration has applied a novel technique to the analysis of 
two particle azimuthal correlations which  extinguishes the harmonic part  of the 
underlying event revealing the true jet shape. Recent extensions of the method to 
three particle correlations allow for a more revealing study of jet topologies 
in Au+Au collisions at ($\sqrt{s_{_{\rm  NN}}}$=200~GeV).}


\PACS{see: http://www.aip.org/pacs/ }
 
\makeindex
\begin{document}
 
\maketitle

\section{Introduction}

The  energy density achieved in Au + Au collisions at RHIC far exceeds the lattice QCD 
estimate for creating the QGP. The high matter density gives rise to large pressure 
gradients which are the driving force for the observed large azimuthal anisotropy ($v_2$) 
of particle emission from the collision zone. The value of this anisotropy is close 
to the predictions of the hydrodynamic model which in turn implies the creation of a 
strongly interacting medium which undergoes early thermalization \cite{RLacey_QM05}. 
Jets provide good probes of this medium provided one can decompose 
the jet signal from the collective flow effects. Possible medium associated 
modifications of the jet topology are a conical emission due to a ``sonic boom" 
effect \cite{Casalderrey_04} and deflection induced by interactions with the partonic 
flow \cite{Armesto_04}. Two and three particle azimuthal correlations can be an 
effective tool in the study of jet topology. 

\section{Two Particle Azimuthal  Correlations}

To study jet topologies we use two- and three-particle azimuthal correlation functions.
For two-particle correlations, the correlation function 
$C\left( {\Delta \phi } \right)$ is given by
$
C\left( {\Delta \phi } \right)=\frac{N_{real} \left( {\Delta \phi } 
\right)}{N_{mix} \left( {\Delta \phi } \right)},
$
where ${\Delta \phi }$ is the difference of the azimuthal angles of the pair. The 
real distribution ($N_{real}\left( {\Delta \phi }\right)$) is built from pair members 
belonging to the same event and the mixed distribution ($N_{mix}\left( {\Delta \phi }\right)$) 
is made of pair members belonging to different events. Thus the correlation function is free 
of geometric acceptance effects and carries only the combined correlations from flow and jets. 
Decomposition of these correlations into their jet and flow contributions, constitute an 
important prerequisite for obtaining the jet function and hence, information about jet 
fragmentation.

\begin{figure}[!htb]
\vskip -0.2cm
\begin{minipage}{0.5\linewidth}
\includegraphics[width=1.\linewidth]{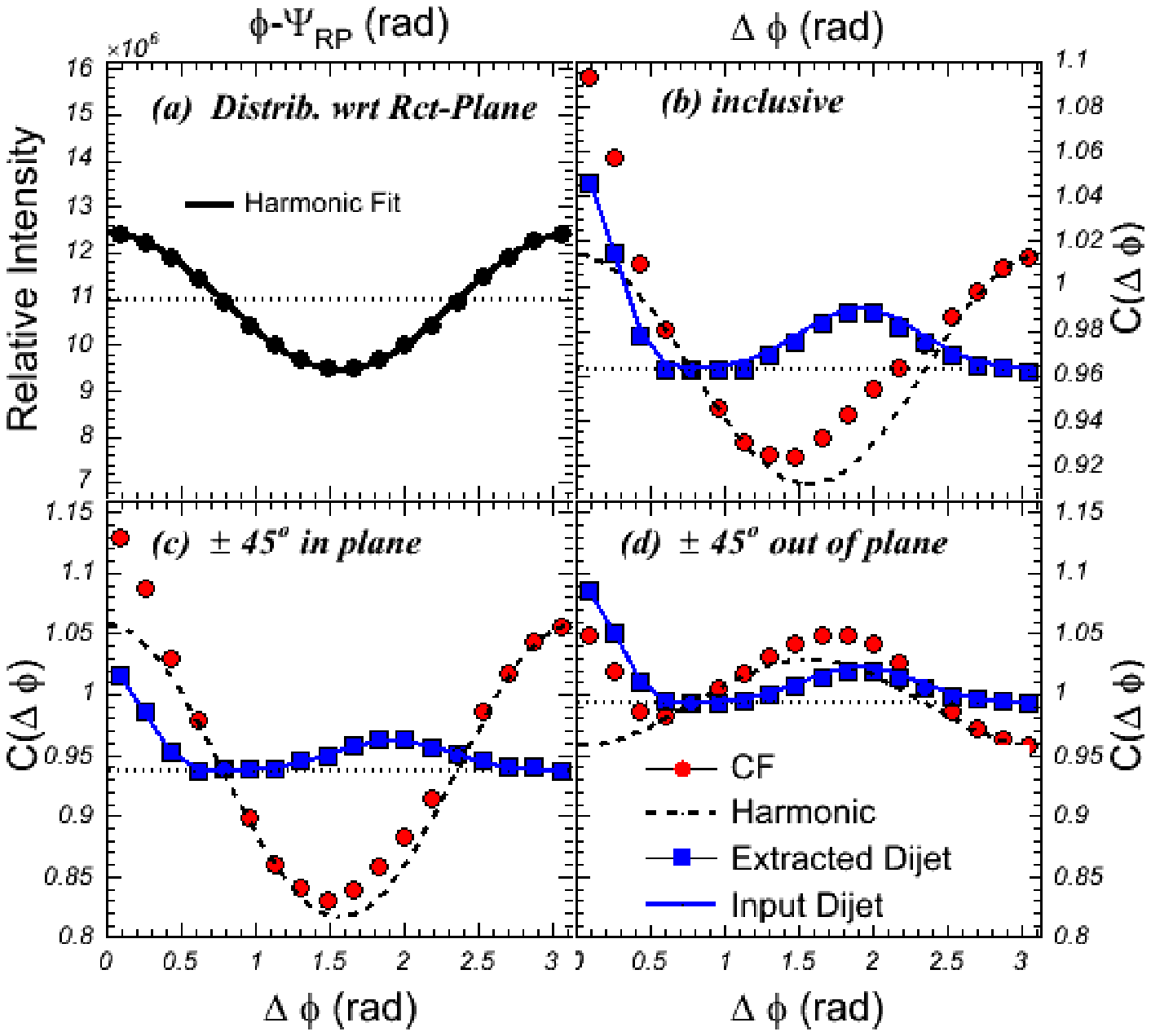}
\vskip -0.7cm
\caption{Simulated data for strongly distorted away-side jets. 
The squares show retrieved points compared to the solid curves for the input jet functions.} 
\label{decomposition1}
\end{minipage}
\hskip 0.1cm
\begin{minipage}{0.5\linewidth}
\includegraphics[width=1.\linewidth]{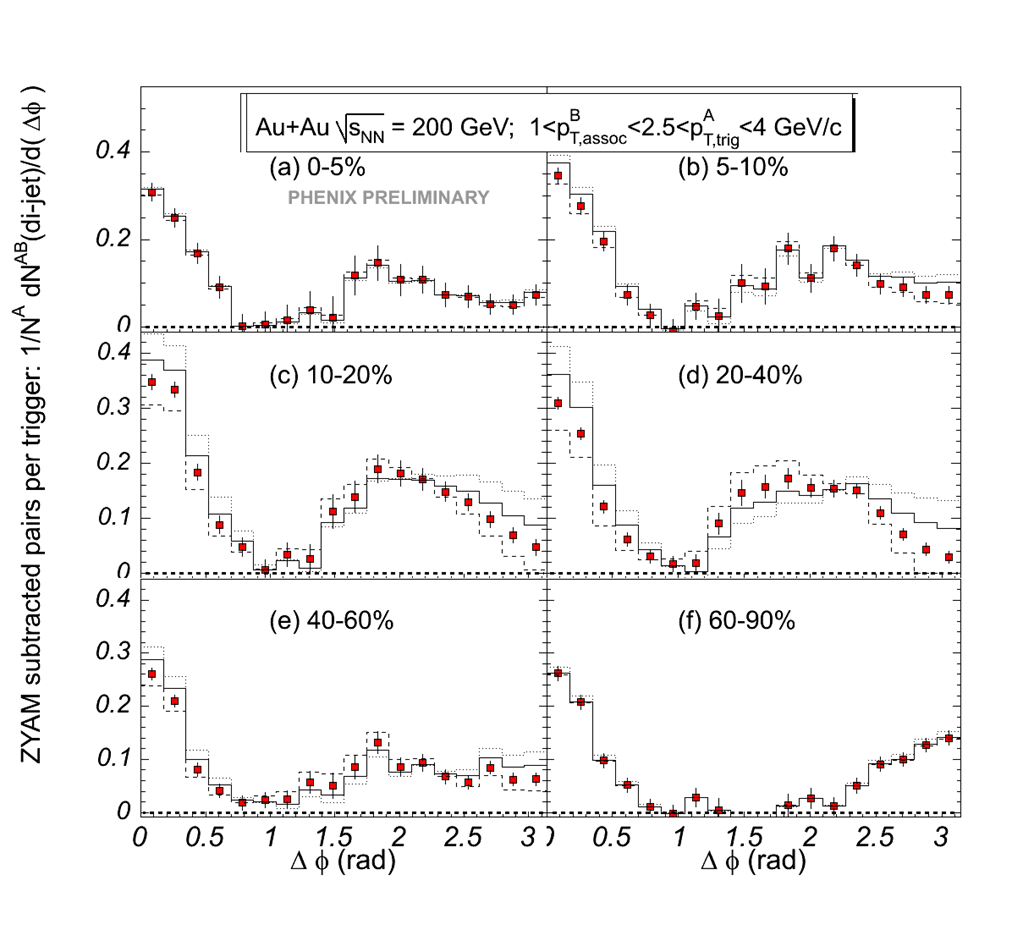}
\vskip -0.7cm
\caption{\small{Experimental jet-pair distributions [after flow subtraction] for several 
centrality selections. }} 
%
\label{ppg32}
\end{minipage}
\vskip -0.5cm
\end{figure}

Figure~\ref{decomposition1} show results  from simulations in which strongly distorted 
away-side jets were studied. Panels (b), (c), and (d) show cases for inclusive, in-plane
and out-of-plane correlation functions. The figure clearly shows that our decomposition 
method retrieves the input jet function in detail, confirming that the decomposition 
procedure is robust even for unusual di-jet distributions.
Figure~\ref{ppg32} show results obtained from the decomposition of the two-particle 
correlation measurements. The apparent shape distortions of the away-side jet is consistent with
recent conjectures of a strong coupling between such jets and the high energy density matter
that they traverse \cite{Casalderrey_04,Armesto_04}. 
\section{Three Particle Azimuthal Correlations}

	Three-particle correlation functions consisting of a trigger hadron from 
the range $2.5< p_T <$4.0 GeV/c (hadron \#1) and two associated hadrons from 
the range $1.0< p_T <$2.5 GeV/c (hadron \#2 and \#3) were also studied (cf. Fig.~\ref{coord_sys}).
Correlation surfaces were constructed by way of $\Delta\phi_{1,2}$ 
and $\Delta\phi_{1,3}$ distributions. 
The correlation surface 
$
C\left( {\Delta \phi_{1,2}, \Delta\phi_{1,3}} \right)=\frac{N_{real} \left( {\Delta \phi_{1,2}, \Delta\phi_{1,3}} 
\right)}{N_{mix} \left( {\Delta \phi_{1,2}, \Delta\phi_{1,3}} \right)}.
$
Here $\Delta \phi_{1,2}$ and $\Delta\phi_{1,3}$ are the azimuthal angle 
difference between trigger and associated particle pairs. The mixed distributions 
were made of pair members belonging to different events. 
Consequently, the correlation functions contain both triples and doubles contributions.


%
One can extinguish the harmonic contributions to the correlation function by aligning 
the high $p_T$ particle perpendicular to the reaction plane followed by adjustment 
of a constraint ``byte-angle" $\phi_c$, to achieve extinction \cite{Ajitanand_05}; 

%
\[
v_2^{out}(\mathrm{trig}) =\left( {\frac{2v_2 \left( {\Delta \phi _c } \right)-\sin \left( 
{2\Delta \phi _c } \right)\left\langle {\cos \left( {2\Delta \Psi _R } 
\right)} \right\rangle +\frac{v_2 }{2}\sin \left( {4\Delta \phi _c } 
\right)\left\langle {\cos \left( {4\Delta \Psi _R } \right)} \right\rangle 
}{2\left( {\Delta \phi _c } \right)-2v_2 \sin \left( {2\Delta \phi _c } 
\right)\left\langle {\cos \left( {2\Delta \Psi _R } \right)} \right\rangle 
}} \right).
\]
%
%

\section{Results} 
%

\begin{figure}[!htb]
\vskip -0.4cm
\begin{minipage}{0.31\linewidth}
\includegraphics[width=1.\linewidth]{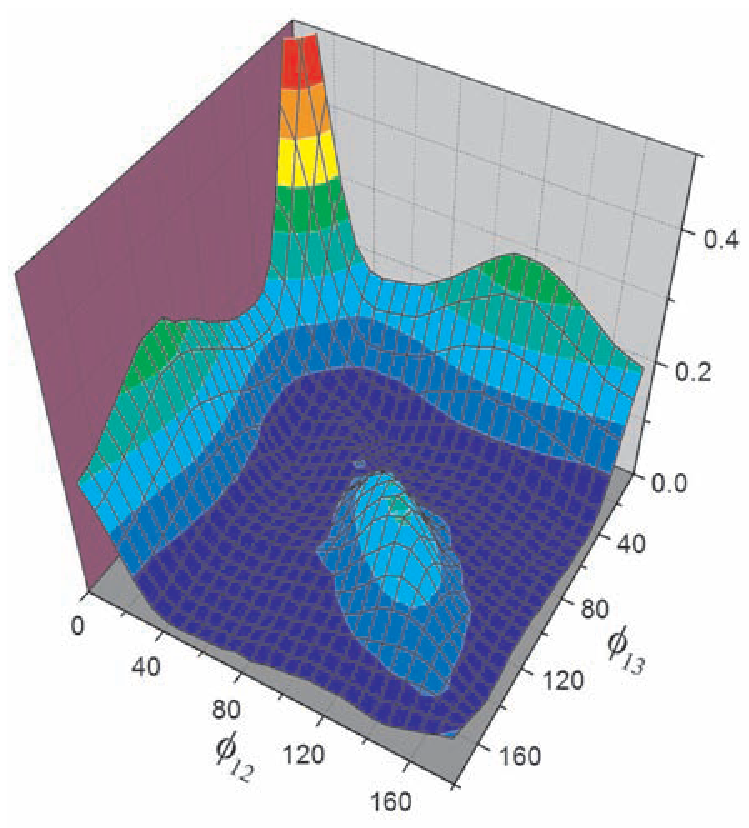}
\vskip -0.8cm
\caption{\small{ Simulated 3-particle correlations for ``deflected" jets.}}
\label{3pc_sim_bjet}
\end{minipage}
\hskip 0.15cm
\begin{minipage}{0.31\linewidth}
\includegraphics[width=1.\linewidth]{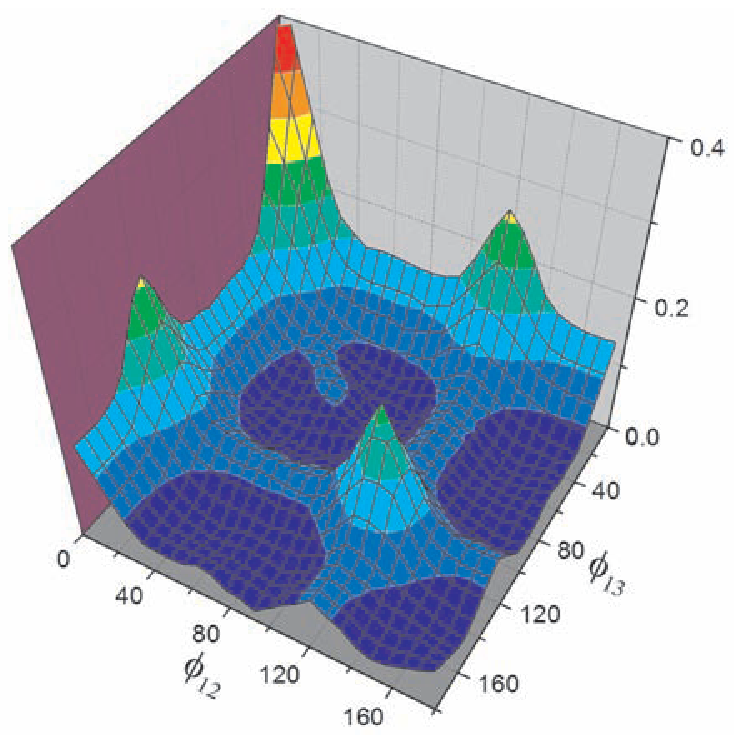}
\vskip -0.7cm
\caption{\small{ Simulated 3-particle correlations for conical flow.}}
\label{3pc_sim_cjet}
\end{minipage}
\hskip 0.15cm
\begin{minipage}{0.31\linewidth}
\includegraphics[width=1.\linewidth]{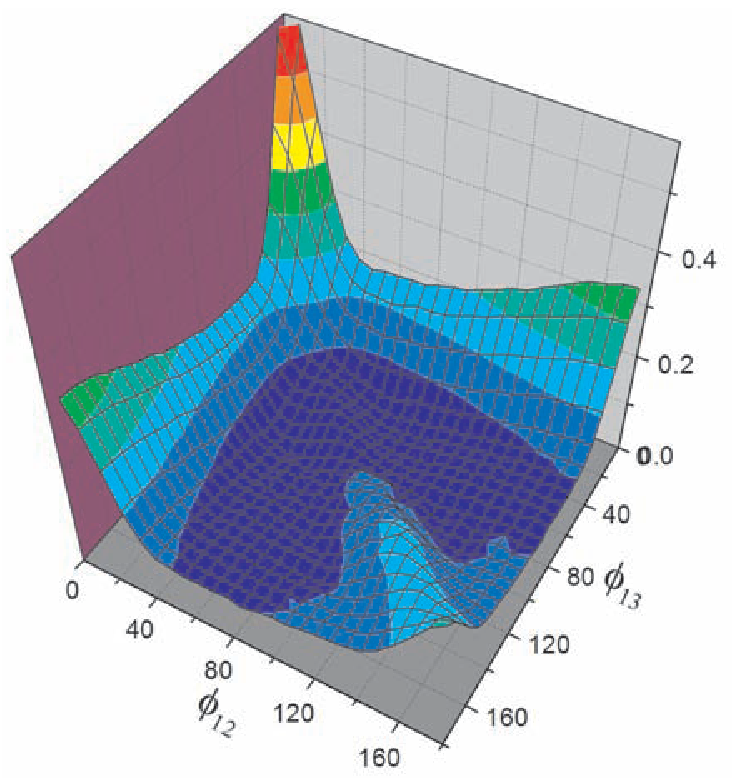}
\vskip -0.8cm
\caption{\small{ Simulated 3-particle correlations for ``normal" jets.}}
\label{3pc_sim_njet}
\end{minipage}
%
\vskip -0.7cm
\end{figure}

Simulated three-particle correlation surfaces 
($\Delta\phi_{1,2}$ vs. $\Delta\phi_{1,3}$) are shown in 
Figs.~\ref{3pc_sim_bjet}~-~\ref{3pc_sim_njet} for three distinct away-side jet scenarios; 
(i) a ``normal jet" in which the away-side jet axis is aligned with the leading jet axis with a spread, 
(ii) a ``deflected jet" in which the away-side jet axis is misaligned by $\sim 60^o$, and 
(iii) a ``Cherenkov or conical jet" in which the leading and away-side jet axes are 
aligned but fragmentation is confined to a very thin hollow cone with a half angle 
of $\sim 60^o$. The simulated results show relatively clear distinguishing features 
for the three scenarios considered.

\begin{figure}[!htb]
\vskip -0.5cm
\begin{minipage}{0.31\linewidth}
\includegraphics[width=1.\linewidth]{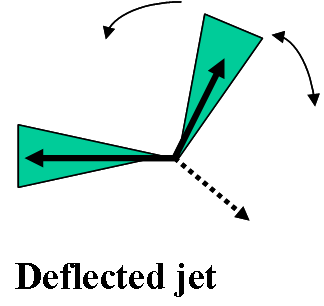}
\vskip -1.0cm
\end{minipage}
\hskip 0.15cm
\begin{minipage}{0.31\linewidth}
\includegraphics[width=1.\linewidth]{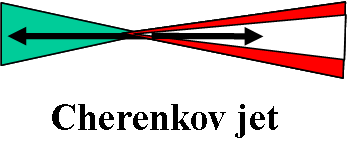}
\vskip -1.0cm
\end{minipage}
\hskip 0.15cm
\begin{minipage}{0.31\linewidth}
\includegraphics[width=1.\linewidth]{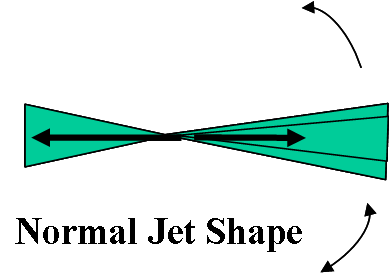}
\vskip -1.0cm
\end{minipage}
%
\end{figure}
\begin{figure}[!htb]
\begin{minipage}{0.31\linewidth}
\includegraphics[width=1.\linewidth]{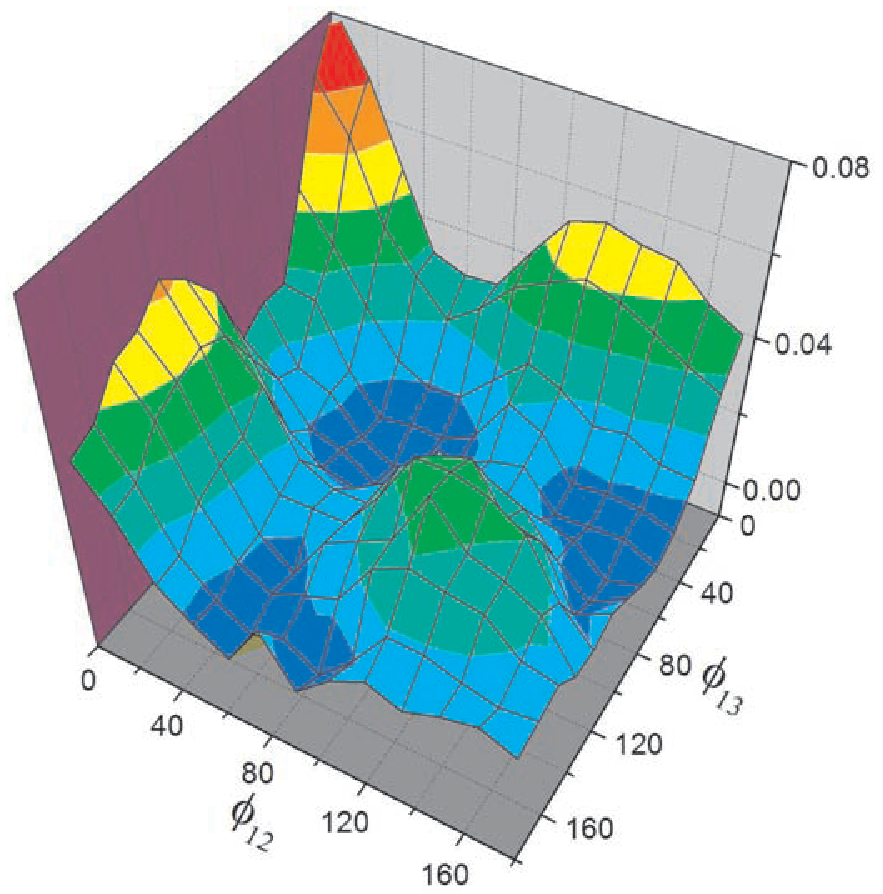}
\vskip -0.7cm
\caption{\small{ Hadron-hadron-hadron correlation function.}}
\label{3pc_data_hhh}
\end{minipage}
\hskip 0.2cm
\begin{minipage}{0.31\linewidth}
\includegraphics[width=1.\linewidth]{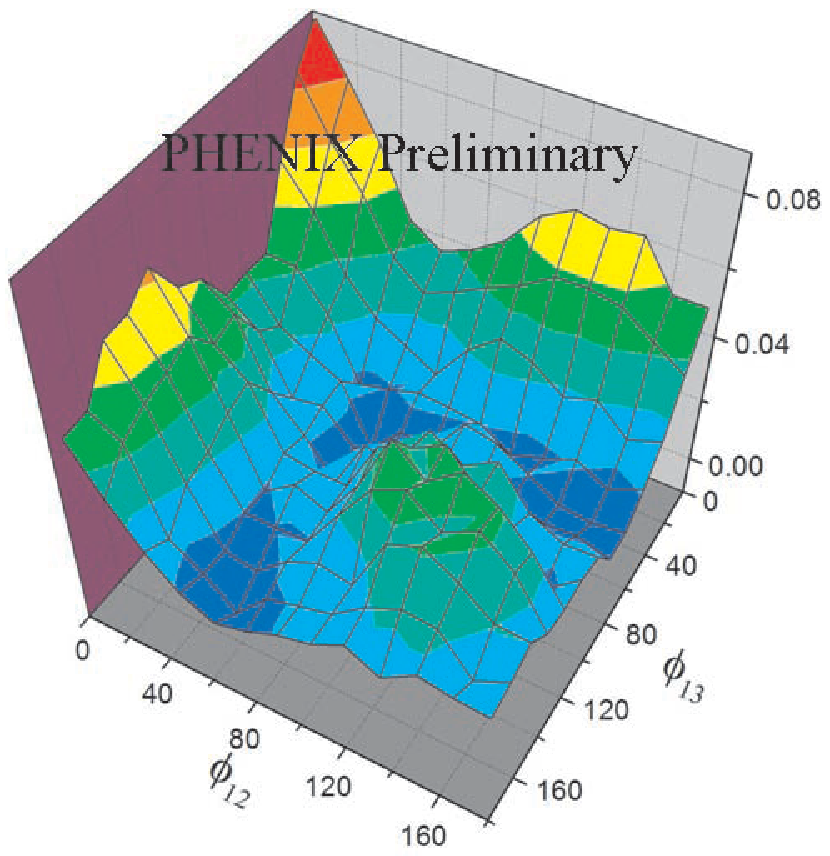}
\vskip -0.7cm
\caption{\small{ Hadron-meson-meson correlation function.}}
\label{3pc_data_hmm}
\end{minipage}
\hskip 0.2cm
\begin{minipage}{0.31\linewidth}
\includegraphics[width=1.\linewidth]{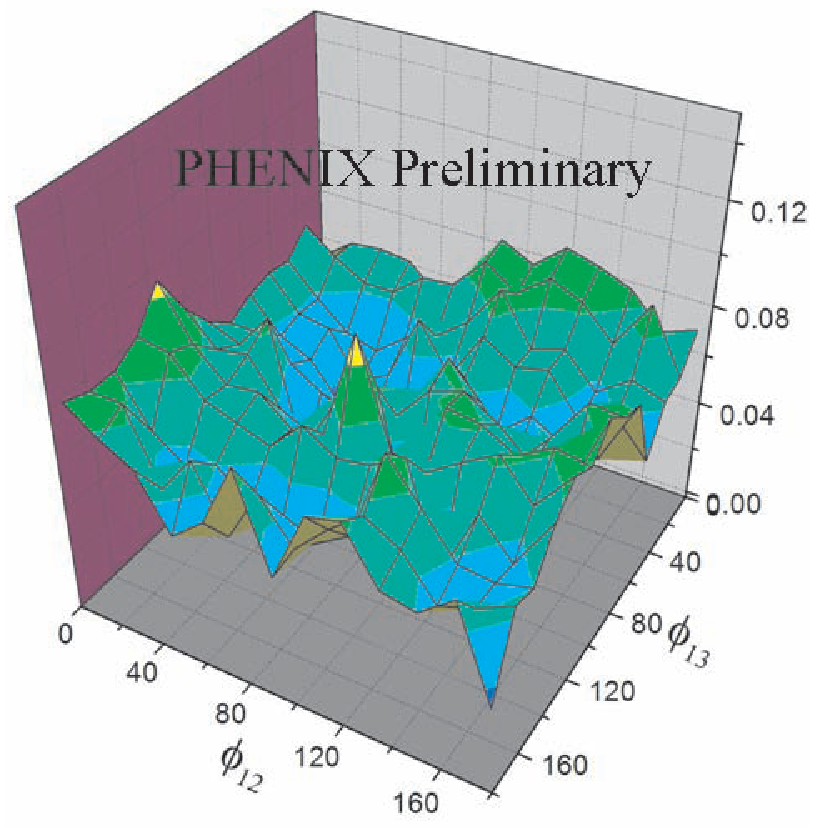}
\vskip -0.7cm
\caption{\small{ Hadron-baryon-baryon correlation function.}}
\label{3pc_data_hbb}
\end{minipage}
\hskip 0.2cm
\begin{minipage}{0.31\linewidth}
\end{minipage}
%
\vskip -0.6cm
\end{figure}
%
The correlation surfaces obtained from data for the centrality selection 10-20 {\%} are shown 
in Figs.~\ref{3pc_data_hhh}~-~\ref{3pc_data_hbb}. They show a strong dependence on the flavor (PID) 
of the associated particle and clearly do not follow the expected patterns for a ``normal jet". 
We conclude that these three-particle correlation surfaces provide additional compelling evidence 
for strong modification of the away-side jet.
Further detailed quantitative investigations are however required to firm up 
the signatures in the data which distinguish between a ``deflected jet" and 
a ``Cherenkov jet".

\section{Summary} 

Two and three particle correlation functions have  been analyzed to 
extract the away-side jet structure for a high $p_T$ trigger-particle in 
association with low $p_T$ particles. Jet landscapes and jet-pair distributions 
have been obtained as a function of event centrality and particle flavor. 
Preliminary comparisons to simple simulations do not exclude "Mach cone" like 
features in the data.


\begin{thebibliography}{9}
%
%
\bibitem{RLacey_QM05} R. Lacey, nucl-ex/0510029.
%
\bibitem{Casalderrey_04} Casalderrey-Solana,Shuryak,Teaney hep-ph/0411315
\bibitem{Armesto_04} Armesto, Salgado, Wiedemann, hep-ph/0411341
\bibitem{Ajitanand_05} Ajitanand Alexander, Chung, Holzmann, Issah, Lacey, Shevel, Taranenko P. 
                  Danielewicz  Phys.Rev. C 72, 011902 (2005)
%
\bibitem{Bielcikova_04} Bielcikova,Esumi,Filimonov,Voloshin,Wurm Phys. Rev. C 69, 021901 (2004)

\bibitem{Phenix_et} K. Adcox et al, PRL 87, 052301 (2001).
%
\end{thebibliography}
\end{document}